\documentclass[a4paper,twoside]{article}

\usepackage{epsfig}
\usepackage{subfigure}
\usepackage{calc}
\usepackage{amssymb}
\usepackage{amstext}
\usepackage{amsmath}
\usepackage{amsthm}
\usepackage{multicol}
\usepackage{pslatex}
\usepackage{apalike}
\usepackage{SciTePress}
\usepackage{graphicx}
\usepackage{morefloats}
\usepackage[small]{caption}

\begin{document}

\title{Non-uniform Quantization of Detail Components in Wavelet Tranformed Image for Lossy JPEG2000 Compression. }

\author{\authorname{Madhur Srivastava\sup{1} and Prasanta K. Panigrahi\sup{2}}
\affiliation{\sup{1}Cornell Center for Technology Enterprise and Commercialization, Cornell University, Ithaca, New York, USA.}
\affiliation{\sup{2}Department of Physical Sciences, Indian Institute of Science Education and Research, Mohanpur, West Bengal, India.}
\email{ms2736@cornell.edu, pprasanta@iiserkol.ac.in}
}


\abstract{The paper introduces the idea of non-uniform quantization in the detail components of wavelet transformed image. It argues that most of the coefficients of horizontal, vertical and diagonal components lie near to zeros and the coefficients representing large differences are few at the extreme ends of histogram. Therefore, this paper advocates need for variable step size quantization scheme which preserves the edge information at the edge of histogram and removes redundancy with the minimal number of quantized values. To support the idea, preliminary results are provided using a non-uniform quantization algorithm. We believe that successful implementation of non-uniform quantization in detail components in JPEG-2000 still image standard will improve image quality and compression efficiency with lesser number of quantized values.}

\onecolumn \maketitle \normalsize \vfill

\section{\uppercase{Introduction}}
\label{sec:introduction}

\noindent The emergence of JPEG (Joint Photographic Experts Group)-2000 still image compression standard has led to a different approach to image analysis of image data compared to the previous JPEG standard \cite{Marcellin2000}. It has higher compression efficiency and better error resilience. Furthermore, it is progressive by resolution and quality in comparison to other still image compression standards \cite{Skodras2001}. These new features have enabled the use of JPEG-2000 in new areas like internet, color facsimile, printing, scanning, digital photography, remote sensing, mobile, medical imagery, and e-commerce \cite{Skodras2001}. One of the major reasons for better performance and the wide range of applications of JPEG-2000 is due to the introduction of Discrete Wavelet Transform (DWT) in the standard replacing Discrete Cosine Transform. In addition, the quantization of DWT coefficients of image has led to rate-distortion feature in the standard. Part-I of JPEG-2000 image compression standard uses fixed-size uniform scalar deadzone quantization scheme. In Part-II, deadzone with variable length is incorporated in uniform scalar quantization. Additionally, trellis coded quantization (TCQ) scheme is combined in the Part-II of JPEG compression standard \cite{Marcellin2002}. However, there are some limitations on the step size selection in the case of lossy image compression using non-invertible wavelets (see section 3).

This paper advocates the need for non-uniform quantization scheme in the detail components of DWT for lossy image compression to overcome the current limitations, to improve image quality at same size, and to improve the compression ratio at a particular image quality. The main motivation for advocating non-uniform quantization is that in each detail component, the majority of the coefficients lie near to zero and the coefficients representing large differences are few at the extreme ends of histogram. This is due to the fact that each detail component represents the high frequency coefficients and in an image, high frequency coefficients only occur at the edges which constitute extremely low percentage of the entire image. Therefore, there is a need for a quantization scheme that provides variable step size with bigger step size around the zero, and smaller step size at the ends of the histogram plot of each horizontal, vertical and diagonal component. On the other hand, the approximation component is untouched with non-uniform quantization. This is because in approximation component almost all the frequency coeffients have substantial magnitude and contain high amount of information, unlike detail components which have high redundancy. 

The paper is organized in the following way. Section 2 briefly describes discrete wavelet transform. In section 3, uniform quantization scheme is given with its limitations on non-invertible wavelets. Section 4 provides a prospective non-uniform quantization algorithm. Preliminary results supporting the proposed algorithm are given in section 6. Finally, section 7 concludes the paper stating the advantages of using the proposed non-uniform quantization algorithm and provides direction for future work.
%\cite{Smith98}.

\section{\uppercase{Discrete Wavelet Transform}}
\noindent In discrete wavelet transform the image is decomposed into four pieces at each level. It is identical to the subbands spaced in the frequency domain. At first level the original image is decomposed into four levels which are labeled as LL, LH, HL and HH as shown 
\begin{figure}[th]
  %\vspace{-0.2cm}
  \centering
   {\epsfig{file = 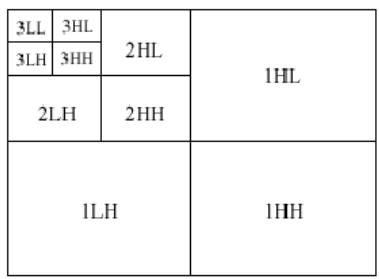, width = 5cm}}
  \caption{Representation of three level 2-D DWT.}
  \label{fig:example1}
 \end{figure}
\begin{figure}[th]
  %\vspace{-0.2cm}
  \centering
   {\epsfig{file = 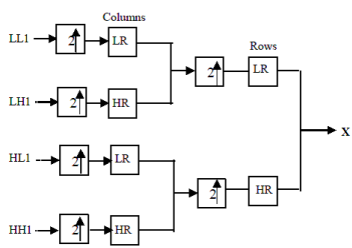, width = 5cm}}
  \caption{Wavelet filter bank for one level image.}
  \label{fig:example1}
 \end{figure}
in fig. 1. Where LL subband is called the average part. It is low pass filtered in both the directions and it is most likely identical to the original image, hence it is called approximation. LH is the difference of the horizontal rows, HL gives the vertical differences and HH gives the diagonal difference. The components LH, HL and HH are called detailed components. The approximation part is further decomposed until the final coefficient is left. The decomposed image can be reconstructed using a reconstruction filter. Fig. 2 shows the wavelet filter bank for one level image reconstruction. Since approximation part is identical to original image hence it contains wavelet coefficients of larger amplitudes. On the other hand in detailed components, wavelet coefficients are smaller in amplitude and are close to zero.

\section{\uppercase{Uniform Scalar Quantization}}
\noindent In Part-I of the standard, JPEG-2000 applies deadzone uniform scalar quantizater on the wavelet transformed image. The quantization index \textit{q} is calculated using the following formula,

\begin{equation}\label{eq1}
   q=sign(y) \lfloor \frac{|y|}{\Delta} \rfloor
\end{equation}

where $\Delta$ is quantizer step size and \textit{y} is the input to the quantizer.

The uniform scalar quantizer with variable size of deadzone modifies the above formula to the following,

\begin{equation}
q= \left\{
\begin{array}{l l}
0 & |y| < -nz\Delta \\
sign(y) \lfloor\frac{|y| + nz\Delta}{\Delta}\rfloor & |y| \ge -nz\Delta \\
\end{array} \right.
\end{equation}

It has to be noted from equation 2 that the fixed width of step size is maintained, except for the deadzone region.
However, in the case of non-invertible wavelets there are some limitations regarding step size and its binary representation \cite{Marcellin2002}. Firstly, there has to be only one quantization step size per subband. This constraints the step size to be less than or equal to the quantization step size of all the different regions of interest in the subband. Secondly, the step size can acquire upto 12 significant binary digits. Thirdly, the upper bound of step size can quantize almost all the coefficients of subband to zero when the step size is twice or more than the subband magnitude. Lastly, the lower bound on the step size restricts high accuracy coding as upto 21 and 30 fractional bits of HH coefficients are allowed to be encoded for 8 and higher bits, respectively.

\section{\uppercase{Proposed Non-Uniform Quantization Algorithm}}

\noindent An algorithm is proposed for the quantization of each detail component in the wavelet domain into variable step sizes using mean and standard deviation. Starting from the weighted mean of histogram plot, the algorithm is recursively applied on the sub-ranges to find the next threshold level and continuing it till ends of the histogram are reached. 
\begin{figure*}[th]
  %\vspace{-0.2cm}
  \centering
   {\epsfig{file = 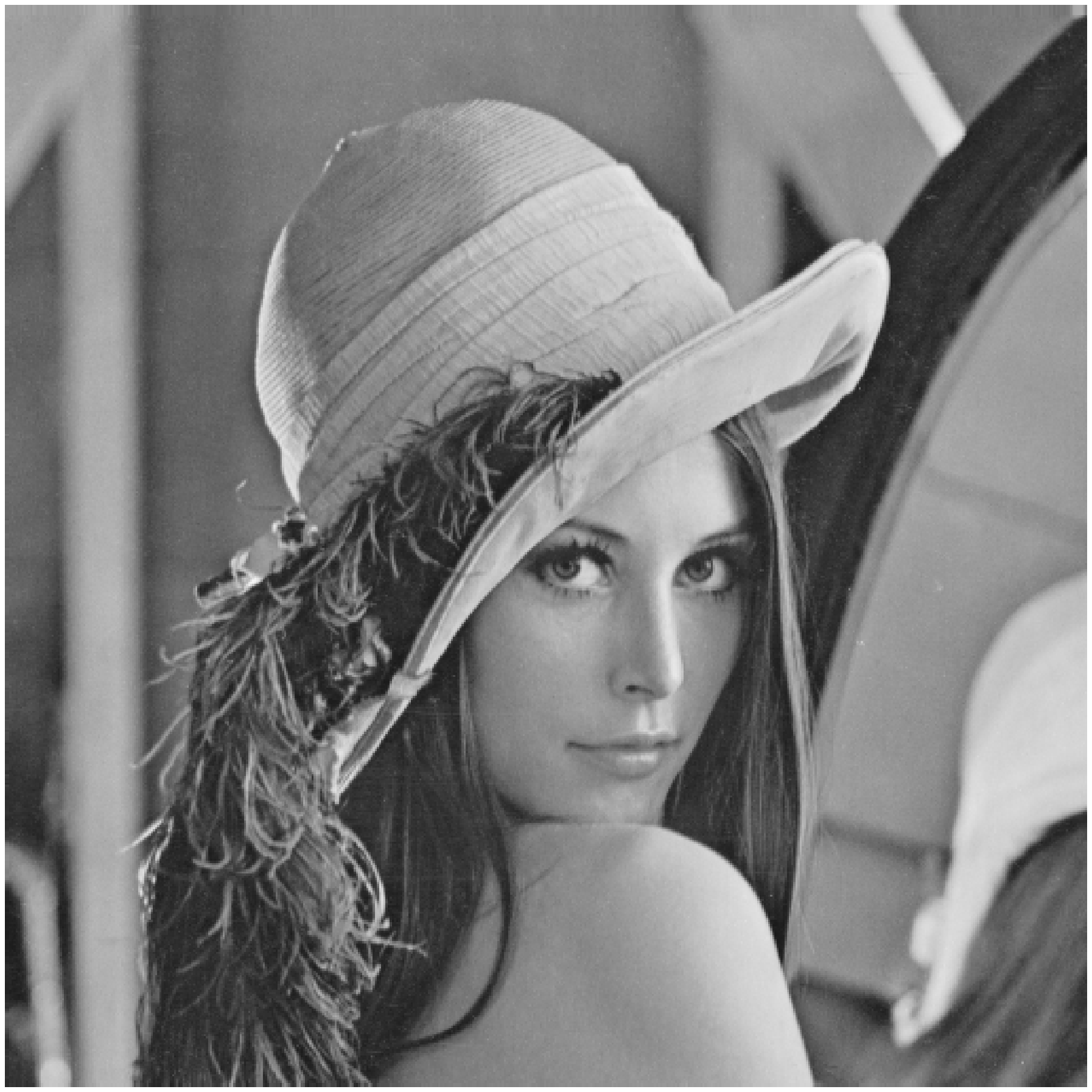, width = 5cm}\epsfig{file = 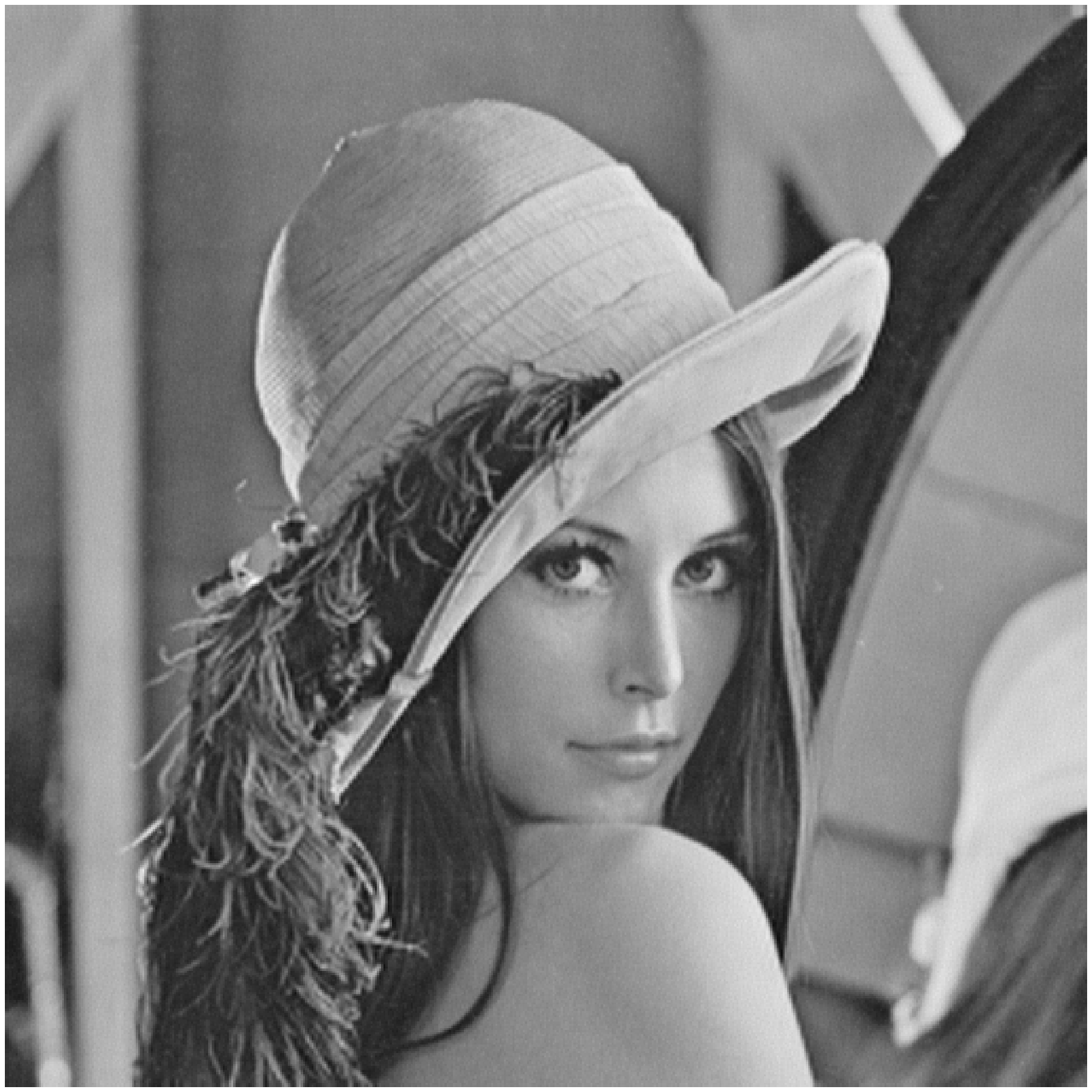, width = 5cm}\epsfig{file = 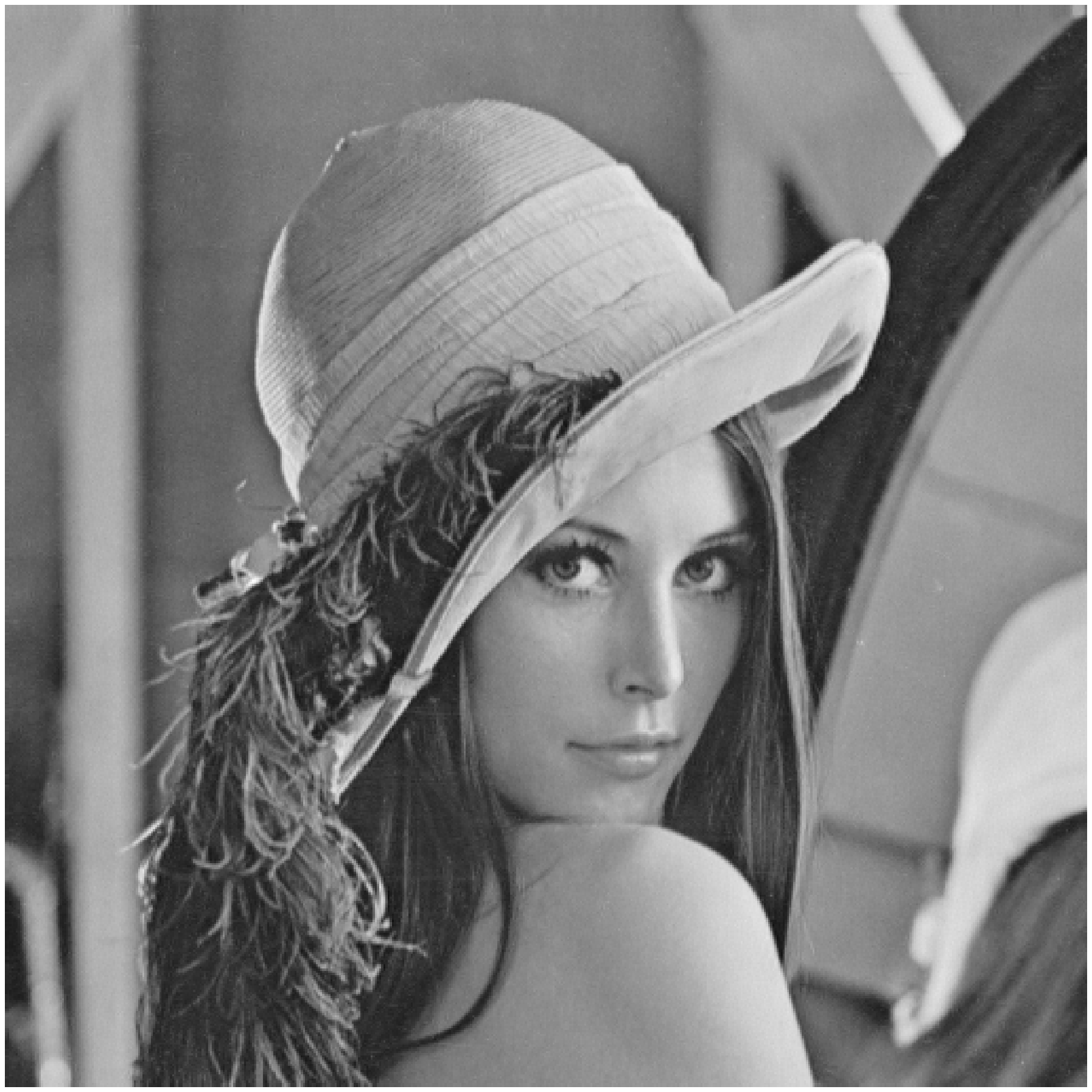, width = 5cm}\\ \epsfig{file = 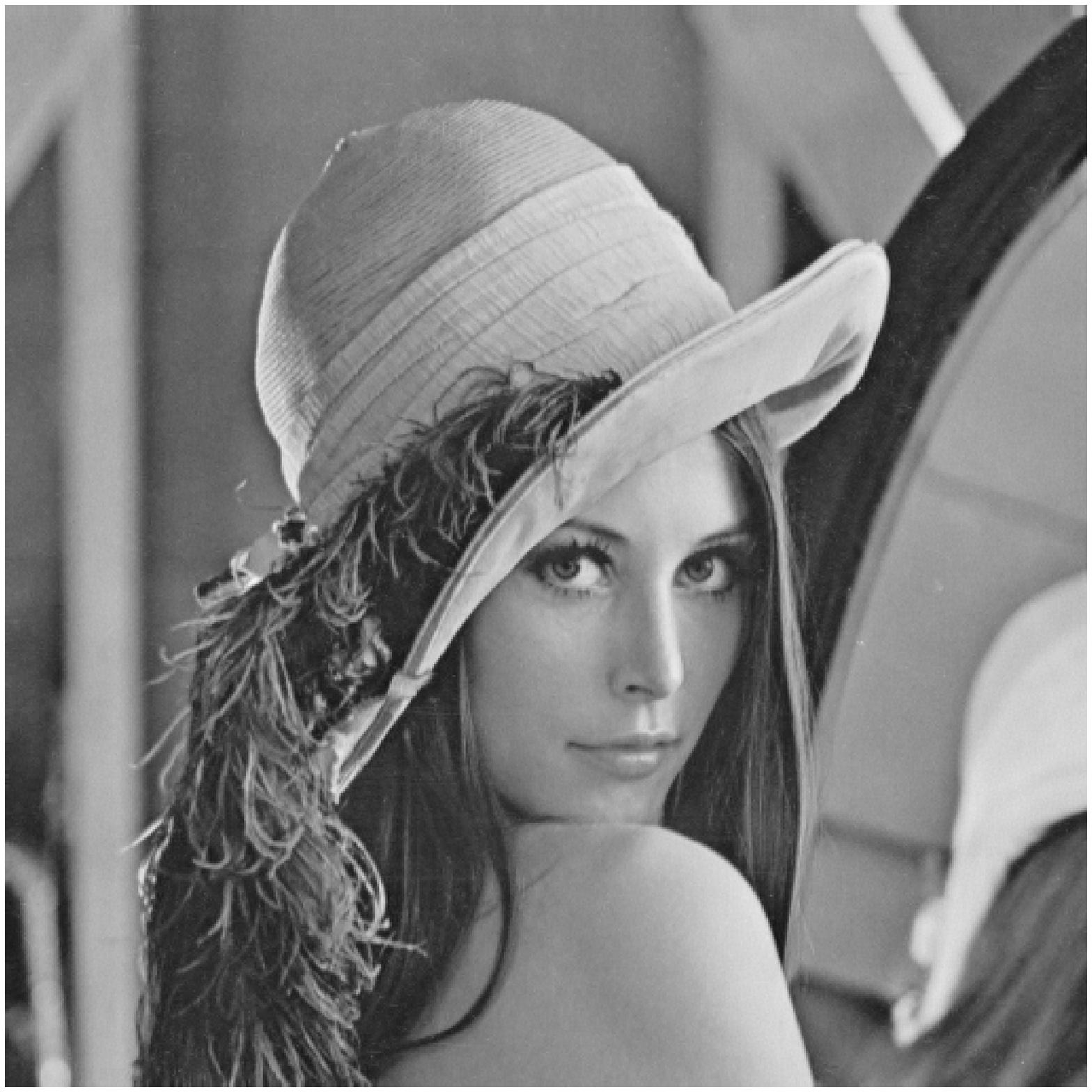, width = 5cm}\epsfig{file = 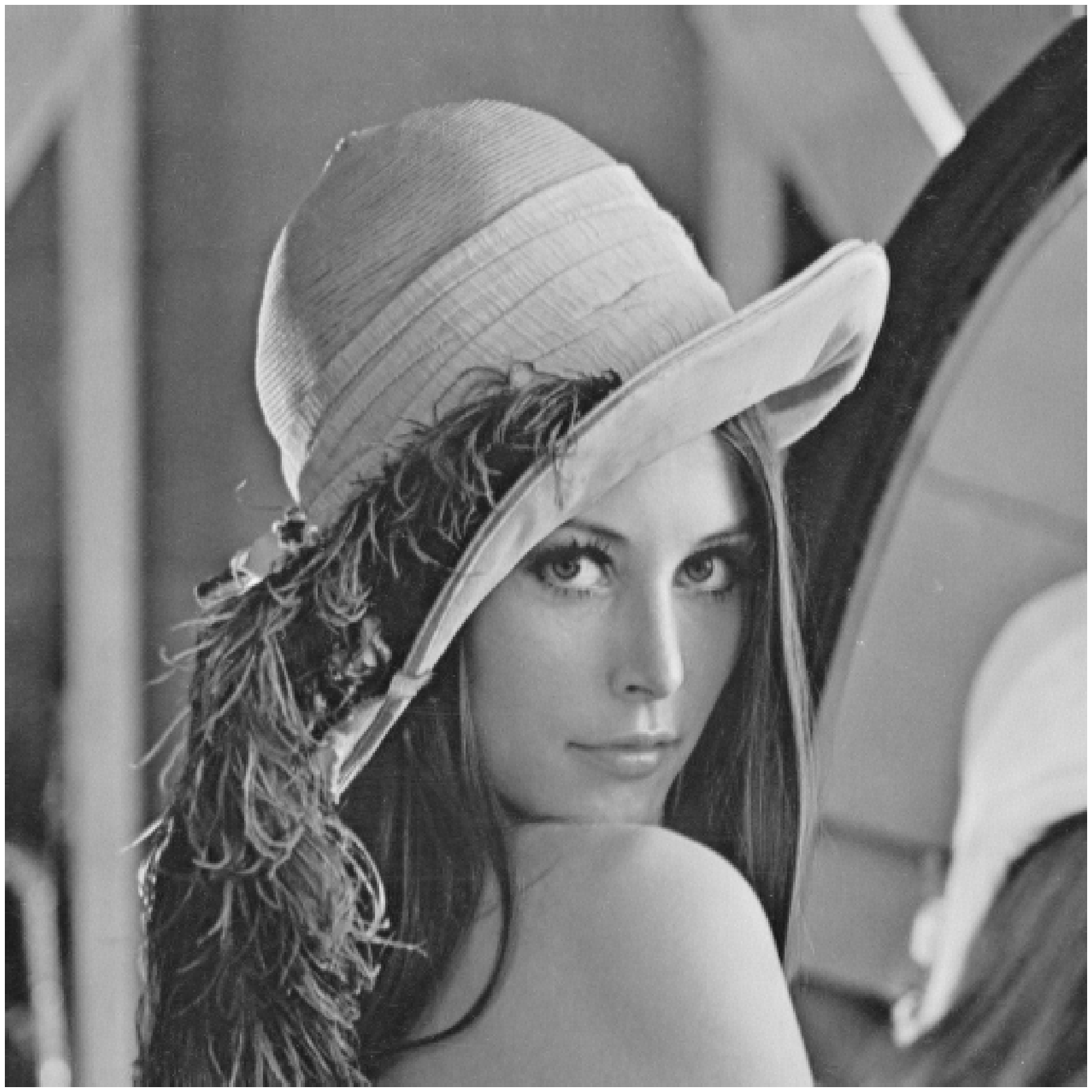, width = 5cm}\epsfig{file = 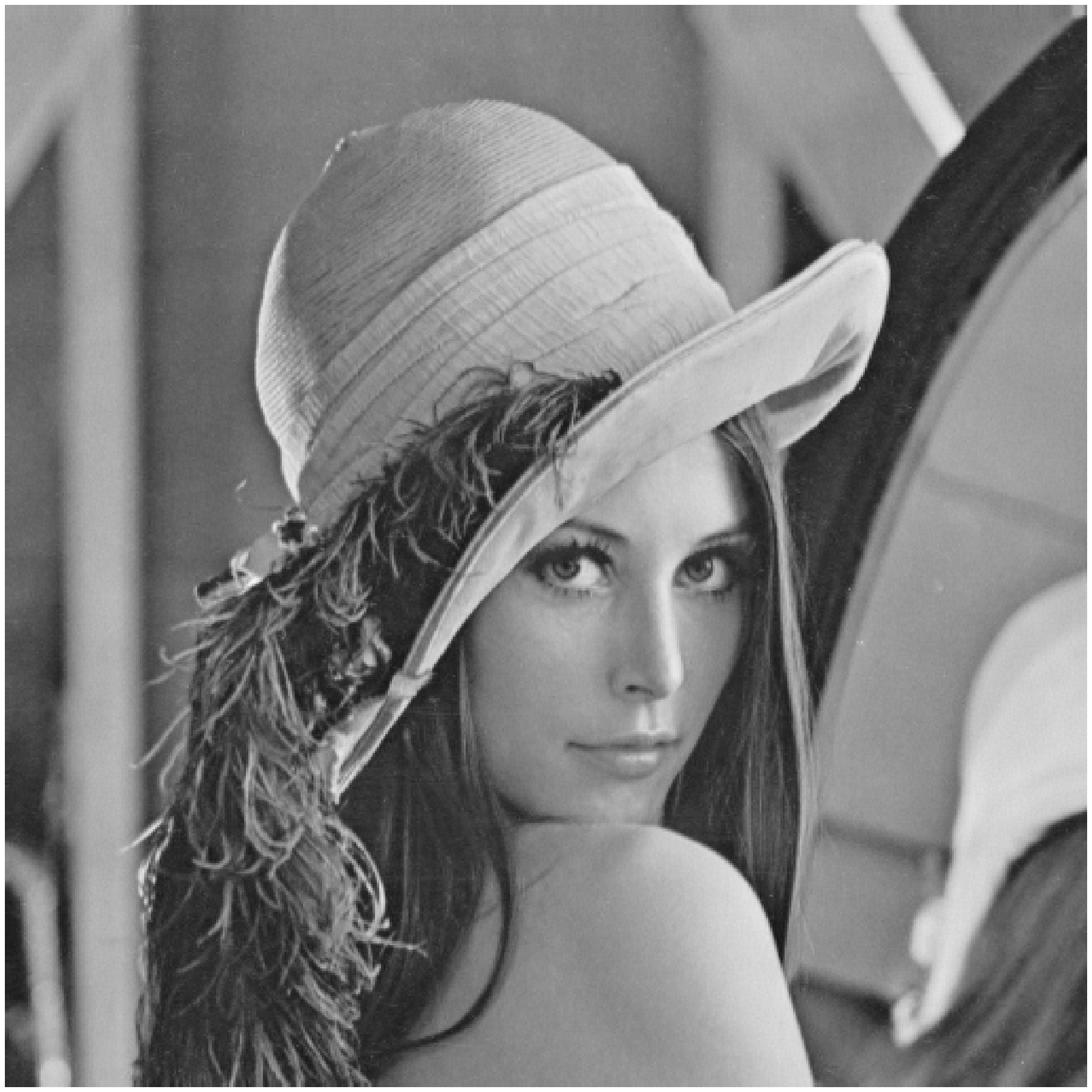, width = 5cm}}
  \caption{Clockwise: Original 'Lenna' image; Reconstructed image at quantized level 2; Reconstructed image at quantized level 4; Reconstructed image at quantized level 10; Reconstructed image at quantized level 8; Reconstructed image at quantized level 6. DB9 wavelets are used in all forward and inverse wavelet transforms.}
  \label{fig:example1}
 \end{figure*}
The method takes into account that majority of coefficients lie near to zero and coefficients representing large differences are few at the extreme ends of histogram. Hence, the procedure provides for variable step size with bigger block size around the mean, and having  smaller blocks at the ends of histogram plot of each horizontal, vertical and diagonal components, leaving approximation coefficients unchanged. The algorithm is based on the fact that a number of distributions tends toward a delta function in the limit of vanishing variance. In the following we systematically outline the algorithm.

\begin{enumerate}
\item $n$, the number of step sizes are taken as input.

\item Range $R=[a,b]$; initially $a=min (histogram)$ and $b=max (histogram)$.
 
\item Find weighted mean ($\mu$) of values ranging in $R$.

\item Initially, thresholds $T_1$= $\mu$ and $T_2$= $\mu$+0.001 . For $T_2$, 0.001 is added to avoid the use of weighted mean again. Any value can be taken such that it doesn’t move significantly from weighted mean.

\item Repeat steps 6-9 $(n-2)/2$ times

\item Find weighted mean ($\mu_1$)  and standard deviation ($\sigma_1$) of values ranging $[a,T_1]$ and weighted mean ($\mu_2$)  and standard deviation ($\sigma_2$) of values ranging $[T_2,b]$.

\item Thresholds $t_1$ and $t_2$ are calculated as $t_1$=$\mu_1$ - $k_1\sigma_1$ and $t_2$=$\mu_2$+ $k_2\sigma_2$ where $k_1$ and $k_2$ are free parameters. These are used to increase or decrease the block size.

\item Assign the values ranging $[t_1,T_1]$ and $[T_2,t_2]$ with their respective weighted mean. 

\item Assign $T_1$=$t_1$-0.001 and $T_2$=$t_2$+0.001 . The value 0.001 is added and subtracted to avoid the reuse of $t_1$ and $t_2$. Any value can be taken such that it doesn’t move significantly from $t_1$ and $t_2$.

\item Finally, Take weighted mean of values ranging $[a,T_1]$ and $[T_2,b]$ and assign the same to the respected range.
\end{enumerate}
%\cite{Moore99}.

The above algorithm was used to compare Daubechies and Coiflet wavelet family on there efficacy to provide effective image segmentation in \cite{Srivastava2011}. Moreover, it is also applied in \cite{SrivastavaUnpublished} to carry out wavelet based image segmentation. However, no study has been conducted of its effectiveness on lossy image compression. \begin{table*}[ht]
\caption{PSNR and MSSIM of reconstructed gray-level test images using DB9 wavelets at different quantized values.}\label{tab:example1} \centering
\begin{tabular}{|c|c|c|c|}
  \hline
Image Name (Dimension) & Number of Quantized Values & PSNR (dB) & MSSIM \\
  \hline
  Lenna (512 $\times$ 512) & 2 &	37.7780645697 &	0.9982142147 \\
& 4 &	41.8874641847 &	0.9992823393 \\
& 6 &	43.8590907980 & 0.9994968461 \\
& 8 &	44.2931530130 &	0.9995311420  \\
& 10 &	44.3489820362 &	0.9995364013  \\
  \hline
  Baboon (512 $\times$ 512) & 2 &	26.8132232589 &	0.9595975231 \\
& 4 &	31.5997241598 &	0.9869355763 \\
& 6 &	32.9223693072 &	0.9905018802 \\
& 8 &	33.0622299131 &	0.9908045725 \\
& 10 &	33.0762467334 &	0.9908358964 \\
  \hline
Pepper (512 $\times$ 512) & 2 &	35.1261558539 &	0.9968488432 \\
& 4 &	38.0511064803 &	0.9982568820 \\
& 6 &	40.6144660901 &	0.9990231546 \\
& 8 &	40.9936699886 &	0.9991423921 \\
& 10 &	41.0082447784 &	0.9991441908 \\
  \hline			
House (512 $\times$ 512) & 2 & 31.9886146645 &	0.9904885967 \\
& 4 & 35.7235201624 & 0.9951268912 \\
& 6 & 38.2730683421 & 0.9975855578 \\
& 8 & 38.4299731739 & 0.9976918850 \\
& 10 & 38.4349224964 &	0.9976960112 \\

  \hline
\end{tabular}
\end{table*}
The preliminary results shown in the next section suggests that the above algorithm can be useful in carrying out non-uniform quantization in detail coefficients.

\section{\uppercase{preliminary Experimental Results}}

\noindent The proposed algorithm is applied on the standard test images from the USC-SIPI Image Database (sipi.usc.edu/database). Peak Signal to Noise Ratio (PSNR) and Mean Structural Similarity Index Measure (MSSIM) \cite{Wang2004} are used to evaluate the quality of reconstructed images. The PSNR is computed using the following formula,

\begin{equation}
PSNR = 10\log_{10} (\frac{I(x,y)^2_{max}}{MSE})
\end{equation}

\begin{equation}
MSE = \frac{1}{MN} \sum_{x=1}^{M} \sum_{y=1}^{N} (I(x,y) - \tilde{I}(x,y))^2
\end{equation}

where $M$ and $N$ are dimensions of the image, $x$ and $y$ are pixel locations, $I$ is the input image, and $\tilde{I}$ is the reconstructed image. The MSSIM is calculated with the help of following formulae,

\begin{equation}
SSIM(p,q)=\frac{(2\mu_p\mu_q + K_1)(2\sigma_{pq} + K_2)}{(\mu_p^2 + \mu_q^2 + K_1)(\sigma_p^2 + \sigma_q^2 + K_2)}
\end{equation}

\begin{equation}
MSSIM(I,\tilde{I})=\frac{1}{M} \sum_{i=1}^{M} SSIM(p_i,q_i)
\end{equation}

where $\mu$ is mean; $\sigma$ is standard deviation; $p$ and $q$ are window sizes of original and reconstructed images, and the size of typical window is $8 \times 8$; $K_1$ and $K_2$ are the constants with $K_1=0.01$ and $K_2=0.03$; M is the total number of windows.

Fig. 3 displays the reconstructed images at different quantized values after applying the proposed algorithm on each detailed component, separately. In addition to the fig. 3, table 1 shows the PSNR and MSSIM of four test images with dimensions $512 \times 512$ at quantized values ranging from 2 to 10. DB9 is the wavelet used for taking DWT and inverse DWT. As can be seen from fig.3, the images are not distinguishable visually. Results from table 1 substantiates the claim. For 'Lenna', minimum and maximum PSNR observed are 37.77 and 44.38, respectively, with MSSIM being .99 at all quantized numbers. Overall, there is increase in PSNR and MSSIM with the increase of the number of quantized values. This increase may or may not be effective for practical application. For example, 'Baboon' will be best represented with 4 quantized levels because there is substantial increase of PSNR and MSSIM from quantized level 2, but compared to quantized 6 to 10, the increase of these higher levels are not effective enough to be noticed visually.

\section{\uppercase{Conclusion}}
\label{sec:conclusion}

\noindent This paper introduces the concept of non-uniform quantization for detailed components in the JPEG-2000 still image standard. We believe that further exploring this option might lead to the following progressive results in the current standard:

\begin{enumerate}
\item Improved quality at same image size.
\item Better compression at same quality.
\item Flexible number of quantized values based on the actual statistics of wavelet tranformed image.
\item Variable step size compared to fixed step size, maximizing the elimination of redundancy.
\end{enumerate}

However, the results shown in the paper are preliminary and suggestive. It will be interesting to examine the proposed quantization algorithm when embedded in the JPEG-2000 standard. Factors which will determine the success of proposed approach will be actual compressed size, image quality, time complexity and encoder complexity. Also, there is wide scope for testing other non-uniform quantization schemes, or perhaps creating new quantization scheme customized to the standard.

%\vfill
\bibliographystyle{apalike}
{\small
\bibliography{example}}

\vfill
\end{document}